Enhancing the Hybrid Software Cost Modeling Method CoBRA® for Supporting Process Maturation

**Supporting Process Maturation with the Enhanced CoBRA® Method**

*Adam Trendowicz, Jens Heidrich, Jürgen Münch*

Fraunhofer Institute for Experimental Software Engineering (IESE)

{adam.trendowicz, jens.heidrich, juergen.muench}@iese.fraunhofer.de

***Summary:***

*Cost estimation is a very crucial field for software developing companies. In the context of learning organizations, estimation applicability and accuracy are not the only acceptance criteria. The contribution of an estimation technique to the understanding and maturing of related organizational processes (such as identification of cost and productivity factors, measurement, data validation, model validation, model maintenance) has recently been gaining increasing importance. Yet, most of the proposed cost modeling approaches provide software engineers with hardly any assistance in supporting related processes. Insufficient support is provided for validating created cost models (including underlying data collection processes) or, if valid models are obtained, for applying them to achieve an organization's objectives such as improved productivity or reduced schedule. This paper presents an enhancement of the CoBRA® cost modeling method by systematically including additional quantitative methods into iterative analysis-feedback cycles. Applied at Oki Electric Industry Co., Ltd., Japan, the CoBRA® method contributed to the achievement of the following objectives, including: (1) maturation of existing measurement processes, (2) increased expertise of Oki software project decision makers regarding cost-related software processes, and, finally, (3) reduction of initial estimation error from an initial 120% down to 14%.*

***Keywords***

*Cost Estimation, Quantitative Analysis, Process Improvement, Iterative CoBRA® Model Improvement, Industrial Case Study.*

## 1    Introduction

The two major prerequisites for selecting a certain cost modeling method as the preferred one are its *applicability* in an organization-specific context and its *contribution* to the achievement of organization-specific objectives.

Method applicability might be perceived as a *necessary acceptance criterion,* since it must be satisfied (the method must first be feasible in a certain context) before one may consider its further characteristics (e.g., estimation accuracy). The type, quantity, and quality of data required by a method as well as the cost of its application are example decision criteria to be considered here. In practice, a careful evaluation of applicability is often missing.

On the other hand, the method's contribution to the achievement of an organization's objectives might be, considered as a *sufficient acceptance criterion*, since if it is satisfied, it guarantees the expected benefit from its application. In the context of learning organizations, estimation accuracy is typically not the only important acceptance criterion. A highly acceptable cost modeling method should contribute to a variety of objectives, such as improving cost-related software processes (e.g., requirements/change management), benchmarking projects with respect to development productivity, or increasing cost estimation expertise. Even though method accuracy is the primary goal, a highly acceptable method should provide means to validate its output and identify sources of potential weaknesses (e.g., validity of underlying measurement processes, estimator expertise, etc.).

This paper presents an enhancement of a hybrid cost estimation method called CoBRA®, [1] and the results of its industrial application in the context of a case study at Oki Electric Industry Co., Ltd. [18]. The CoBRA® method was selected as the best candidate with respect to the Oki context and objectives.

We enhanced the CoBRA® method by quantitative analysis and feedback cycles. In consequence, we were able to (1) identify significant cost drivers for more effective inclusion into the measurement program, (2) improve the measurement processes and, in consequence, the quality of collected project data, and (3) better adapt the CoBRA® model to the specific organizational context.

## 2    Related Work

Numerous cost estimation methods have been developed over the last years. Yet, as reported by the most recent market studies, they did not contribute much to the improvement of successful software project planning [17]. One of the potential reasons is that not all proposed methods are equally applicable in every context. Moreover, published empirical studies on cost estimation models do not provide much support to software practitioners in selecting "the best" method, i.e., a method that would work best in the specific context of their software organization.

In principle, organizational context might be defined as an organization's *capabilities* and *objectives*. An organization's capabilities specify the feasibility of a certain estimation method to be applied in a certain context. An organization's objectives specify usefulness of a certain method in terms of its ability to contribute to the achievement of those objectives. Since a certain method first has to be feasible in a certain context before we can consider its usefulness, we may call those two characteristics *necessary* and *sufficient* criteria for selecting the best estimation method (for a certain context), respectively.

---

[1] CoBRA® is a registered trademark of the Fraunhofer Institute for Experimental Software Engineering.

The type, quantity, and quality of data required by a method as well as the cost of its application are examples of necessary criteria to be considered. We may basically distinguish three groups of methods with respect to the type of data they require: *data-based*, *expert-based*, and *hybrid methods*.

Among *data-based methods*, some require past project data in order to build customized models (define-your-own-model approaches), others provide an already defined model, where factors and their relationships are fixed (fixed-model approaches). The major advantage of a fixed-model approach is that it does not require any data from already completed projects. Fixed models are developed for a specific context and are, by definition, only suited for estimating the types of projects for which the fixed model was built. The applicability of such models across various contexts is usually quite limited. In order to improve their performance in a context other than the one it was built for, a large amount of organization-specific project data is required for calibrating the model to a specific application context [12]. In practice, a single organization is seldom capable of gathering the required amount of data, which makes calibration often an unfeasible task in practice. Moreover, even if required data can be gathered, the calibrated model often does not result in satisfactory improvement of performance [6]. In contrast, define-your-own-model approaches use organization-specific project data in order to build a model that fits a certain organizational context.

*Expert-based methods*, on the other hand, do not provide any explicit model and are usually organization specific. Yet, when based on a single expert (so-called "rule-of-thumb" estimation), prediction may be burdened by large bias and, in consequence, low reliability. Structured methods based on several experts [8] provide more reliable estimates, but cost much effort. The data and cost issues are contradictory in principle: the less effort a method requires to build the estimation model, the more measurement data from previous projects is needed[2].

The common characteristic of data- and expert-based approaches is that the quality of their output strongly depends on the quality of the input (either underlying measurement processes or expert expertise). In practice, software organizations move between those two extremes, tempted either by low application costs or low data requirements. In fact, a great majority of organizations that actually use data-based methods do not have a sufficient amount of appropriate (valid, homogeneous, etc.) data as required by such methods. On the other hand, more and more software organizations are attempting to reach level 3 of the Capability Maturity Model (CMM) [4], in which quantitative collection of measurement data plays a significant role. These companies usually already started data collection, but still rely upon expert knowledge. From a logical point of view, *hybrid* methods such as CoBRA® offer a reasonable bias between data requirements and application costs.

---

[2] The more data is available, the less expert involvement is required and the company's effort is reduced. This does not include the effort spent on collecting the data. Yet, collected data might usually be used for multiple purposes (incerased benefit).

Moreover, hybrid methods have numerous indirect benefits. In addition to simple point estimates, the CoBRA® method, for instance, provides means for cost-related risk assessment and project benchmarking [3]. As a define-your-own-model approach, CoBRA® also provides means for goal-oriented improvement of existing measurement processes. The transparent structure of the model and the integrated sensitivity analysis allow for identifying factors with the greatest influence on software productivity and, in consequence, to focus improvement activities on related processes [14]. So the method does not only support continuous improvement of the model itself, but also improves cost-related organizational processes.

As already mentioned, a contribution to the achievement of an organization's objectives might be considered as a *sufficient method acceptance criterion*, especially in the context of learning organizations, where estimation precision is not the primary acceptance criterion any more. In that context, a highly acceptable cost estimation method should contribute to organization-wide improvement efforts, e.g., by supporting constant monitoring of the technical and organizational processes in order to detect weaknesses and react appropriately (including the estimation processes themselves).

Yet, when we look at hundreds of empirical studies on cost modeling published so far (e.g., [2]) the first impression we may get is that the only reasonable criterion for selecting "the best" method is the precision of the estimates it derives. Cost estimation methods are developed and compared with respect to this one criterion. The second impression would be that this criterion is probably not very helpful for selecting "the best" method, because instead of results that converge, at least, the reader often has to face a contradicting outcome of empirical investigations. There are numerous sources of those deviations. Even if performed on the same data, studies vary with respect to data preparation procedures applied [15], configuration of the evaluated method [16], evaluation strategy adapted [16], or evaluation measures applied [9]. Moreover, there is a great deal of risk in adopting to a certain context a method that provided accurate estimates when applied on data from different (often multiple) organizations and from measurement processes of unknown validity [13].

In summary, there exist only very few studies that consider (at least to some extent) the practical applicability of cost estimation methods. Hardly any cost estimation method provides wide applicability and comprehensive support to the achievement of organizational goals in the context of learning organizations [5].

## 3    The CoBRA® Principles

CoBRA® [1] is a hybrid method combining data- and expert-based cost estimation approaches. The CoBRA® method is based on the idea that project costs consist of two basic components: nominal project costs and a cost overhead portion as presented below.

$$Cost = \underbrace{Nominal\ Productivity \cdot Size}_{Nominal\ Cost} + Cost\ Overhead \quad (1)$$

where

$$Cost\ Overhead = \sum_i Multiplier_i(Cost\ Factor_i) \\ + \sum_i \sum_j Multiplier_{ij}(Cost\ Factor_i, Indirect\ Cost\ Factor_j) \quad (2)$$

Nominal cost is the cost spent only on developing a software product of a certain size in the context of a *nominal* project. A nominal project is a hypothetical "ideal" project in a certain environment of an organization (or business unit). It is a project that runs under optimal conditions; i.e., all project characteristics are the best possible ones ("perfect") at the start of the project. For instance, the project objectives are well defined and understood by all staff members and the customer and all key people in the project have appropriate skills to successfully conduct the project. Cost overhead is the additional cost spent on overcoming imperfections of a real project environment such as insufficient skills of the project team. In this case, a certain effort is required to compensate for such a situation, e.g., team training has to be conducted.

In CoBRA®, cost overhead is modeled by a so-called causal model. The causal model consists of factors affecting the costs of projects within a certain context. The causal model is obtained through expert knowledge acquisition (e.g., involving experienced project managers). An example is presented in Figure 3-1. The arrows indicate direct and indirect relationships. A sign (´+´ or ´-´) indicates the way a cost factor contributes to the overall project costs. The ´+´ and ´−´ represent a positive and negative relationship, respectively; that is, if the factor increases or decreases, the project costs will also increase (´+´) or decrease (´−´). For instance, if *Requirements volatility* increases, costs will also increase. One arrow pointing to another one indicates an interaction effect. For example, an interaction exists between *Disciplined requirement management* and *Requirement volatility*. In this case, increased disciplined requirement management compensates for the negative influence of volatile requirements on software costs.

The cost overhead portion resulting from indirect influences is represented by the second component of the sum shown in (2). In general, CoBRA® allows for expressing indirect influences on multiple levels (e.g., influences on *Disciplined requirement management* and influences on influences thereon). However, in practice, it is not recommended for experts to rate all factors due to the increased complexity of the model and the resulting difficulties and efforts. Further details on computing the cost overhead can be found in [2].

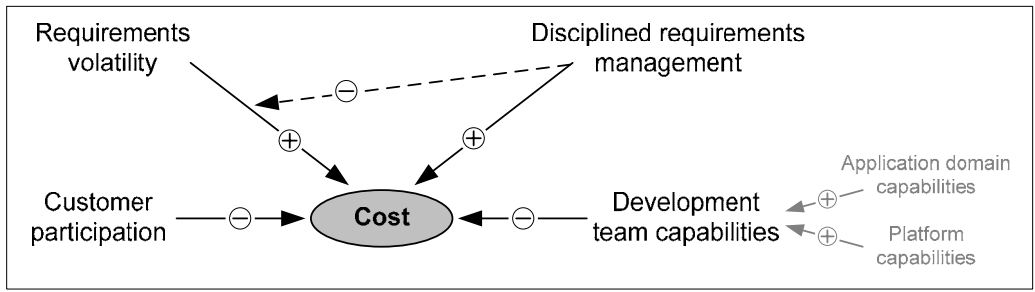

**Figure 3-1:** Causal Model Example

The influence on costs and between different factors is quantified for each factor using expert evaluation. The influence is measured as a relative percentage increase of the costs above the nominal project. For each factor, experts are asked to give the increase of costs when the considered factor has the worst possible value (extreme case) and all other factors have their nominal values. In order to capture the uncertainty of evaluations, experts are asked to give three values: the maximal, minimal, and most likely cost overhead for each factor (triangular distribution).

Based on the quantified causal model and data on past project characteristics, a cost overhead model is generated for each past project using a simulation algorithm (e.g., Monte Carlo or Latin Hypercube). Cost overhead might also be interpreted as productivity loss, and can be used to extract the nominal productivity of each past project (see equation #1). The nominal productivity represents the baseline project productivity without any negative influences of cost factors that makes actual productivity (measured as output/input ratio) differ across projects with various characteristics. In that sense, CoBRA® is also a reliable productivity modeling method. Since equation #1 represents a simple bivariate dependency, it does not require much measurement data. In principle, merely project size and effort are required. The size measure should reflect the overall project volume including all produced artifacts. Common examples include lines of code or Function Points [11]. Past project information on identified cost factors is usually elicited from experts.

The cost overhead probability distribution obtained could be used further to support various project management activities, such as cost estimation, evaluation and benchmarking projects regarding cost-related risks, as well as productivity improvement [1]. Figure 3-2 illustrates two usage scenarios using the cumulative cost distribution: calculating the project costs for a given probability level and computing the probability for exceeding given project costs.

Let us assume (scenario A) that the available budget for a project is 900 Units and that this project's costs are characterized by the distribution in Figure 3-2. There is roughly a 90% probability that the project will overrun this budget. If this probability represents an acceptable risk in a particular context, the project budget may not be approved. On the other hand, let us consider (scenario B) that a project manager wants to minimize the risks of overrunning the budget. In other words, the cost of a software project should be planned so that there is minimal risk of exceeding it. If a project manager sets the maximal tolerable risk of exceeding the budget to 30%, then the planned budget for the project should not be lower than 1170 Units.

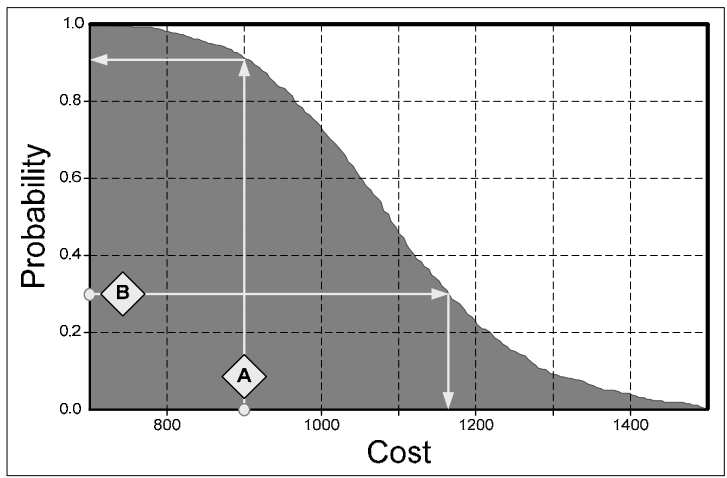

**Figure 3-2:** Example Cumulative Cost Distribution

The advantage of CoBRA® over many other cost estimation methods are its low requirements with respect to measurement data. Moreover, it is not restricted to certain size and cost measures. The method provides the means to develop an estimation model that is tailored to a certain organization's context, thus increasing model applicability and performance (estimation accuracy, consistency, etc.) A more detailed description of the CoBRA® method can be found in [1].

## 4      Shortcomings of the CoBRA® Method

The past industrial applications of CoBRA® proved its usefulness in the context of low availability of measurement data [1]. The cost model was obtained at a relatively low cost, and besides accurate estimates, provided software decision makers with a reliable basis for project risk management. In addition, cost drivers identified within the model served as the basis for building up an organizational measurement program for the purpose of quantitative cost/productivity modeling.

Yet, although CoBRA® contributes to the development of a project measurement database, it does not really benefit from it. Its core element, the cost overhead model, is based exclusively on experts' assessments, which might be a potential source of threats. As we observed during previous CoBRA® applications experts tend to disagree with respect to the most significant cost drivers selected, their relationships as well as their impact on cost. In consequence, the reliability of the cost model largely depends on the experts' cost-related expertise and their understanding of the CoBRA® process. This might be especially true in the context of additional difficulties in the communication between CoBRA® experts and project managers introducing the method (e.g., language, culture, terminology, etc.).

## 5     The Enhanced CoBRA® Method

The enhanced CoBRA® method should thus use all available measurement data to support experts in building a cost overhead model by providing them with guidelines on the one hand and feedback on the other hand. Fortifying the CoBRA® method with quantitative methods would benefit from growing organizational maturity (more quantitative data) on the one hand, and contribute to that maturity (better understanding of software processes, higher quality, goal-oriented measurement data), on the other hand. In addition, the method should ensure minimal size and complexity of the cost overhead model with maximal performance (e.g., accuracy and precision).

To achieve these objectives, we proposed the following procedure (Figure 5-1): First, input data is collected and validated against standard criteria such as completeness, consistency, or correctness. Next, in a so-called *pre-modeling* analysis, available measurement data on project characteristics are explored to identify elements of a cost overhead model. One of the so-called *feature selection* techniques [10] might be applied here to identify the most significant cost drivers, followed by a multivariate correlation analysis to identify potential interaction between the extracted cost drivers. Next, the estimation model is built and *post-modeling* analysis is performed in order to evaluate the quality of the model (with respect to specified modeling objectives). For the cost estimation objective, the model's estimation accuracy and precision can be evaluated in a cross-validation (leave-one-out) experiment on past project data. An essential element of the post-modeling analysis is to validate the obtained model not only against its acceptance criteria (e.g., meeting modeling objectives) but also with respect to the performance (results) of the *pre-modeling* analysis. The feedback loop between pre- and post-modeling analysis provides information on improvement potentials with respect to the model, to related organizational processes (e.g., underlying measurement and data collection processes), as well as to the estimation method itself (e.g., pre-modeling techniques used).

In case of unsatisfactory model performance, the respective improvements can be undertaken and the model refinement iteration may be performed. In general, refinement iterations may be repeated until the refinement benefits (e.g., improvement of model performance) are greater than its costs.

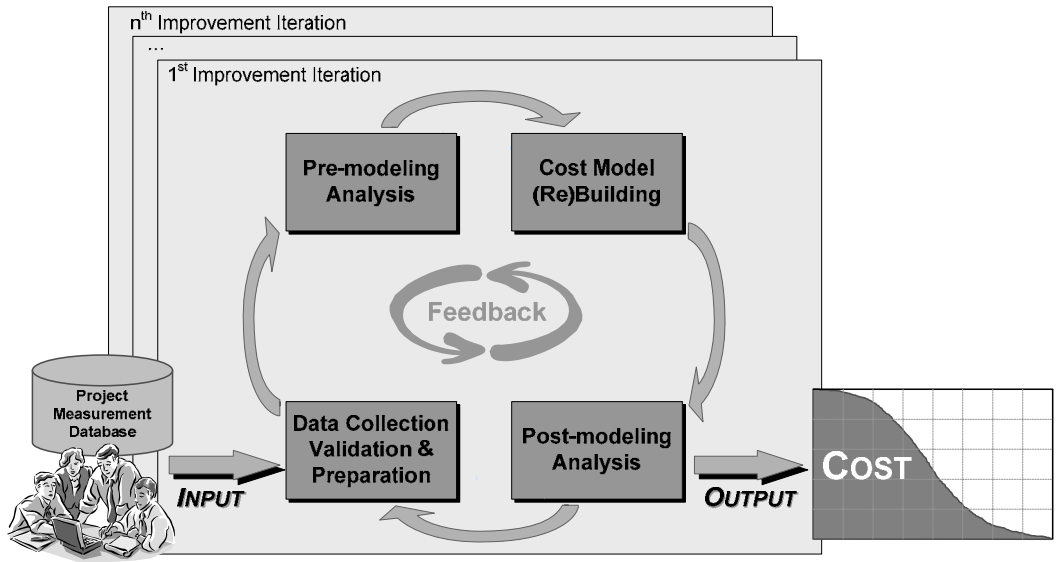

**Figure 5-1:** Enhanced CoBRA Method

## 6 Industrial Case Study

### 6.1 Study context

CoBRA® was applied in the context of an Oki unit dealing with financial applications. Initially, measurement data from 16 past projects was provided and 12 experts (project and quality managers of the company) participated in the study; this means that they assessed past project data and discussed the main cost drivers. All projects were enhancement projects for HP-UX and Windows platforms, and were developed according to a simple waterfall life cycle model. Size was measured as uncommented LOC (for Java and C) and effort as number of person hours.

### 6.2 Improvement iterations

The *initial cost model* was developed using traditional CoBRA® in order to obtain a baseline for comparing the enhanced method. The output model did not, however, meet Oki objectives regarding estimation precision (Figure 6-1, Figure 6-2) and as such could also not be considered a reliable basis for understanding and improving cost-related software processes. Yet, although initial model development revealed certain problems with data consistency (measurements vs. experts' assessments), it was not clear what was the exact source of the model's poor performance. Next, model refinement cycles according to the enhanced CoBRA® method were performed.

During four iterations, several potential improvements of the CoBRA® model as well as related software processes were performed. After removing the project outlier in the 1st iteration, a group of four projects were identified that differentiate significantly from others. The analysis of existing measurement data revealed that the phase of effort measurement is a significant cost diver (makes the considered project differ with respect to productivity). According to Oki's data collection process, effort data was collected correctly for each development phase, but for some projects, the effort spent on requirements specification and/or system testing was not available, and therefore not included in the total effort data.

Such inconsistency of the effort measurement process might have a significant impact on the quality of the CoBRA® model. Therefore, it was decided to modify the respective measurement process (to consistently cover the same project scope for all projects), recollect effort data, and rebuild the CoBRA® model. In principle, there were two possibilities of addressing this issue: (1) include effort data of only those phases of the development life cycle that were consistently measured or (2) recalculate the total normalized effort for the whole life cycle, including the effort for missing phases calculated on the basis of effort distribution across past projects. The latter option of normalizing effort data is, for instance, also applied in the ISBSG database [14]. Normalization, however, would require significant knowledge of historical data with respect to effort distribution across different development phases. Since we did not have sufficient data, we took the first approach.

Further analysis revealed an outlier group of four projects that significantly differ from others regarding productivity. A closer look revealed the use of a second programming language as a significant factor influencing software development cost and differentiating the outstanding group of projects (they were actually partly developed in C, while other projects were completely developed in Java).

Yet, during the joint meeting with Oki experts it was concluded that *Support from project-external technical people* is the factor that makes the outlier projects different from all other projects. The *Use of a second programming language*, identified during the analysis of measurement data was not considered as a crucial cost factor. During further review of the cost overhead model, the involved experts decided moreover, to include several new cost factors. These additional factors include, for instance, *Degree of product enhancement* in order to differentiate between new development and enhancement projects. This was a significant factor differentiating the outlier project removed in the first refinement iteration.

In order to prevent inconsistencies in the expert evaluation, we decided to re-collect past project data for the improved cost overhead model from multiple experts this time. In order to ensure the same understanding of the subjective scales defined to quantify each factor, we invested extra effort into giving a detailed definition of the project situation related to a specific factor value, and also discussed the scales in a group meeting involving all experts.

Re-collection of project data confirmed our initial concerns regarding the reliability of project data elicited from single experts. When acquired from multiple experts, the data showed significant inconsistencies. In several cases, experts gave extremely different evaluations of the same factor in the same project. The problem was solved by a joint experts' meeting where the involved experts discussed the data inconsistencies and came up with a common factor rating. Even though all experts participated and contributed to the detailed definition of the scales for each factor, there were still inconsistencies in interpreting the scales and related project situations.

The refined CoBRA® model showed further improvement in terms of estimation accuracy and precision. Yet, an analysis of measurement data indicated the size of the GUI (Graphical User Interface) and the size of batches as factors having a significant impact on software cost. These factors are missing in the current cost overhead model (i.e., explain the remaining productivity variance not already explained by the current cost overhead model).

After discussing this problem with the experts, it turned out that the currently used size metric reflected only code directly implemented by software developers and did not include other elements of software size (such as the code generated for the GUI and batches). Experts agreed later on that even if, for instance, some parts of the software are generated, they still require a certain effort (e.g., for designing and testing). Thus, the objective of the next refinement iteration was to improve the size metric used, so that it would better reflect the volume of work required to produce project deliverables (empirically valid size measure). Next, respective measurement data had to be recollected, and the CoBRA® model had to be rebuilt.

After refining the size measurement process and recollecting the data, the model performance met the estimation objectives stated by Oki, so it was decided to stop model improvement at this point.

### 6.3 Summary of Model Improvement

Figure 6-1 and Figure 6-2 present a summary of the CoBRA® model application throughout the four refinement iterations. The overview shows constant improvement in estimation accuracy and consistency from one iteration to the next.

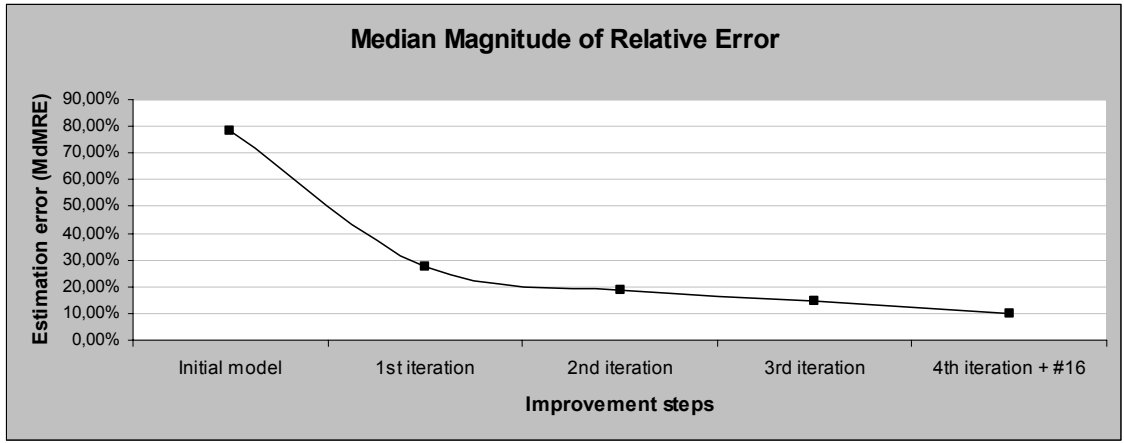

**Figure 6-1:** Improvement of Estimation Accuracy

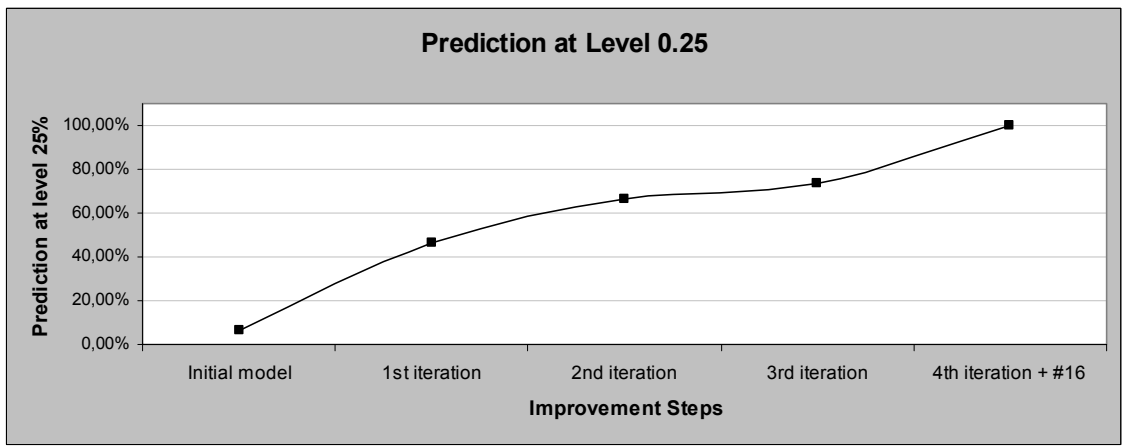

**Figure 6-2:** Improvement of Estimation Consistency

In addition, numerous improvements in estimation-related processes were introduced. The measurement procedures were adjusted to consistently collect effort data for a uniquely defined project scope, and the size metric used was modified in order to fully quantify the size of the software development product that contributes to the overall project effort. Moreover, factors included in the cost overhead model that are not already measured should be included in the organization's measurement program. In addition, experts suggested the cost overhead model should be a part of the organization's experience base [1], for instance, as a means for sharing cost-related knowledge with new employees.

## 7    Summary and Discussion

In this paper, we presented an enhancement of the CoBRA® cost modeling method. We modified the method by supporting experts with additional quantitative methods within an iterative analysis-feedback cycle while building the cost model.

Enhancing the Hybrid Software Cost Modeling Method CoBRA® for Supporting Process Maturation

Applied in an industrial study, the CoBRA® method contributed to the achievement of the software organization's objectives, including: (1) maturation of existing measurement processes and, in consequence, collected project data; (2) increased expertise of project and process managers regarding cost-related software processes; and, finally, (3) reduction of initial estimation error from an initial 120% down to 14%.

Moreover, we gained evidence that, fortified with additional quantitative methods the enhanced CoBRA® performs well in the context of a Japanese software organization where additional communication problems proved to have a negative impact on the applicability and usefulness of the traditional CoBRA® method, which is based mainly on experts.

Due to the relative small size of the data analyzed in the study, we limited our analysis to simple statistical methods such as, correlation-based feature selection analysis. Future work should therefore focus on more robust quantitative methods that may be applied to analyze larger datasets. Three major areas of analysis should be (1) identification of significant cost drivers, (2) analysis of relationships between identified cost drivers, and (3) quantification of the impact of different factors on cost.

## Acknowledgements

We would like to thank Oki Electric Industry Co., Ltd., where we conducted the study and applied CoBRA®, as well as all involved Oki experts and local organizers, who greatly contributed to the successful performance of the project. We would also like to thank the Japanese Information-technology Promotion Agency (IPA) for their support in conducting the study.